\documentclass[reprint,onecolumn]{revtex4-2} 

\usepackage{hyperref,verbatim, physics}
\usepackage{braket,lineno,color,soul, amsmath}
\usepackage[inline]{enumitem}
\usepackage{xcolor}

\newcommand{\IM}[1]{\operatorname{Im}\{#1\}}

\begin{document}

\title{Correlation induced out-of-phase plasmon in an electron liquid}
\author{Gabor J. Kalman}
\affiliation{Department of Physics, Boston College, Chestnut Hill, MA 02467, USA}
\author{Kenneth I. Golden}
\affiliation{University of Vermont, Burlington, VT 05401, USA}
\author{Luciano G. Silvestri}
\affiliation{Michigan State University, East Lansing, MI 48823, USA}

\begin{abstract}
We derive from first principles the existence of a low-frequency plasmon in a strongly coupled three-dimensional homogeneous electron gas (HEG). From its dispersion and its satisfaction of the 3\textsuperscript{rd} frequency sum rule we identify it with the conjectured magnetic excitation in the HEG. This excitation is maintained by the out-of-phase oscillations of the spin-up and spin-down densities of the electron liquid, but governed solely by the Coulomb interaction between the particles. The frequency square of this mode is proportional to the overlap ($r=0$) (absolute) value of the spin-up/spin-down correlation function, and thus slightly affected by, but not contingent upon the degree of polarization of the electron liquid. We estimate the spectral weight of the mode: it is expected to be governed by electron-electron collision induced spin drag. The spectral weight is manifest in the partial spin-resolved dynamical structure functions and it is proportional to the product of the densities of the two spin components. An independent derivation based on a generalized Feynman Ansatz corroborates our result. The relationship to the recently identified ``spin plasmon" excitation is discussed.
\end{abstract}

\maketitle
The existence of plasmons in many-body systems interacting through a Coulomb potential (quasi-uniform plasma, homogeneous electron gases (HEG), etc.) with a characteristic oscillation frequency, the plasma frequency $\omega_p = \sqrt{4 \pi Z^2 e^2 n / m}$ (the symbols having their usual meaning), was first observed by Tonks and Langmuir \cite{Tonks1929}. Its theoretical analysis started with the work of Vlasov \cite{Vlasov1938}. Landau's subsequent criticism \cite{Landau1946} of some aspects of Vlasov's work led to a deeper understanding of the wave-particle interaction and the ensuing damping mechanism. The identification of plasma oscillations as a collective excitation is due to the pioneering series of works by Bohm, Gross and Pines \cite{Bohm1949,Bohm1949a,Bohm1953,Pines1952}, who also introduced the Random Phase Approximation (RPA). It was also Bohm and Gross (BG) \cite{Bohm1949,Bohm1949a} who determined the eponymous wave number $k$-dependent positive dispersion of the plasmon, caused by the random motion of the particles. The understanding of the central role plasma oscillations play in the dynamics of the degenerate electron gas arrived through the series of contributions by Pines and Nozieres \cite{Nozieres1958} who developed the theory of plasma oscillations in the electron gas through the application of the RPA (see also\cite{Sawada1957,Sawada1957a,Klimontovich1960}). Soon, however, it became clear that both the classical Vlasov treatment, its quantum equivalents \cite{Lindhard1954,Ehrenreich1959}, the BG collective coordinate technique, and the RPA share a common underlying theoretical foundation and are appropriate for weak coupling only. Here the coupling strength is defined as the ratio of the potential energy of the particles to their kinetic energy. This for quantum systems is $r_s=a/a_B$, (where $a$ is the Wigner-Seitz radius, $a_B$ the Bohr radius) while the classical equivalent is $\Gamma = e^2/(ak_BT)$ (the symbols having their usual meaning). Motivated by the case of the electron gas in metals where $r_s > 1$, it was Singwi and collaborators \cite{Singwi1968} who made the first serious attempts to study the effect of correlations on the properties of the plasmon. However, protected by the Kohn sum rule \cite{Brout1959}, the plasmon is an extremely robust excitation, unaffected by correlations, \textit{i.e.} the plasma frequency is $r_s$ independent. Therefore, all the correlational effects are absorbed by the wave number $k$-dependence of the dispersion. 

The physics of the correlated HEG (or, rather, homogeneous electron \textit{liquid}) has by now been studied through a large body of theoretical (for a recent summary see \cite{Giuliani2008}) and computer simulation works \cite{Ceperley1978,Jones1996,Ortiz1994,Ortiz1997,Gori-Giorgi2000,Gori-Giorgi2002,Zong2002,Spink2013}. It is well understood that the HEG behaves as a binary system consisting of spin-up and spin-down components. In this binary spin liquid due to exchange, pairs of electrons with parallel or anti-parallel spin orientations correlate differently. As a result, the interaction energy of the pair depends on the relative spin orientations. This feature leads to the possibility of different phases of the HEG as in the higher coupling domain the two components separate in momentum space and the system becomes polarized. Computer simulations show that up to $r_s = 40$ the three-dimensional HEG is still paramagnetic; a continuous second-order transition from the fully unpolarized phase to a partially polarized phase is predicted at $r_s = 50 \pm 2$  and the system becomes ferromagnetic near the Wigner lattice transition, around $r_s \cong 100$ \cite{Zong2002}. 

It was realized as early as 1973 that the impact of the two-component character of the HEG must be manifest in the various dynamical characterization of the system. In the classic papers of Goodman and Sjolander \cite{Goodman1973} and Singh and Pathak \cite{Singh1974} it was shown how the 3\textsuperscript{rd} frequency moment sum-rule coefficient is affected by this feature of the HEG. Further works clarified that in addition to the standard density dynamical structure function (DSF) an additional ``magnetic" DSF constructed from the difference between the fluctuations of the up and down spin densities plays a role \cite{Goodman1973,Giuliani2008}. Utsumi and Ichimaru \cite{Utsumi1983} conjectured the possible existence of a longitudinal out-of-phase spin density oscillation (OPO) mode, but came to the (erroneous) conclusion that it cannot exist within a physical range of coupling values. Atwal and Aschroft \cite{Atwal2003} focused attention on the magnetic DSF and based on a phenomenological study concluded that the OPO excitation should exist.

Thus, by now there is a convergence of studies pointing at the likelihood of an OPO mode. There is no conclusive first principles based evidence, though, of the existence of this mode. Moreover, in the absence of an underlying dynamical model, nothing is known about its dispersion characteristics and spectral weight.

It is important to understand that the model Hamiltonian, from which the indications for the existence of the OPO are derived, similarly to that used in this work, does not distinguish between the interactions of parallel and anti-parallel pairs: both are simple Coulomb forces. For the system to exhibit an out-of-phase oscillation, the different correlations play the role of physical markers. For this reason the RPA description of a binary mixture doesn't show the existence of the OPO (see, however, the comment in the summary below) and it is the strongly coupled phase of the HEG where one should look for the excitation of this mode. 

It is well established by molecular dynamics (MD) simulations \cite{Donko2002} and theoretical analysis \cite{Kalman1990,Golden2000} that classical charged particles exhibit quasi-localization in the strongly coupled phase, \textit{i.e.} they are trapped in fluctuating local potential minima, and it is the oscillation of these temporarily caged particles that govern the formation of the collective modes in the strongly coupled domain. It is expected that the strongly coupled HEG behaves in a similar fashion: this is corroborated by recent works by Drummond \textit{et al.} \cite{Drummond2004}.

In this Letter we contend that a recent series of work on the strongly coupled classical binary systems of charged particles (binary ionic mixtures - BIM) \cite{Kalman2014,Silvestri2015} opens a new way to approach the fundamental physics of the OPO. In particular, the Quasi-Localized Charge Approximation (QLCA) \cite{Kalman1990,Golden2000} which was originally developed for strongly coupled classical systems, should provide an appropriate framework for the description of the collective excitations in the HEG as well. The QLCA predictions have been extensively verified by simulations \cite{Ohta2000,Kalman2000}.
The application of the QLCA to quantum systems, though, requires some further considerations. What should remain valid in the quantum regime is that once the particles occupy a set of fairly well-defined quasi-equilibrium positions their dynamics can be described in terms of coupled oscillating dipoles. This philosophy that was first formulated in the Born-von Karman classical harmonic phonon approximation has been carried through to the description quantum crystals as well. Thus its extension to charged liquids, which is the basic idea of the QLCA, should not distinguish between classical and quantum systems either.  
This similarity notwithstanding the quantum dynamics of the degenerate electron liquid exhibits features that are absent in a classical plasma. These effects need to be included in the QLCA formalism in order for it to be applicable to the electron liquid.

Here we have to emphasize that the QLCA addresses a system in the strong coupling regime, where the particles are quasi-localized. For the electron liquid this is a much more precise requirement than for a classical plasma. Whether a classical trajectory exhibits features of localization is a matter of gradation, and is determined by somewhat arbitrary criteria \cite{Donko2002}. In the electron liquid, in contrast, whether the description of the system in terms of free particle momentum states forming a Fermi distribution or in terms of localized Wannier-type wave functions is more appropriate constitutes a clear-cut difference.
 
We describe the system as a binary Coulomb mixture of $N_\uparrow$ spin-up and $N_\downarrow$ spin-down electrons ($N = N_\uparrow + N_\downarrow$ and total density $n$) immersed in a neutralizing uniform background of smeared out positive charges at $T = 0$. As to the dissipative processes in the system, we recall that Landau damping should be negligible at strong coupling values. Focusing, then, on collisional damping, one observes that at $k=0$ the \textit{intraspecies} damping vanishes, due to momentum conservation \cite{DAmico2000}. The damping in this domain is due to \textit{interspecies} collisions, whose effect may be described in terms of a viscous drag force proportional to $ \gamma |\vb v_{\uparrow} (\vb r) - \vb v_{\downarrow}(\vb r)|,$ where $\vb v_\uparrow, \vb v_{\downarrow}$ are the local hydrodynamic velocities of the spin-up, spin-down components and $\gamma$ is the drag coefficient \cite{Silvestri2015}. The soundness of this approach in conjunction with the QLCA method for other systems has been corroborated by a series of computer simulations, which have provided excellent agreement with calculations based on the model. The issue of the viscous damping (``spin drag") has been studied both theoretically \cite{DAmico2010,Takahashi2007,Yashenkin2015}, and proved experimentally \cite{Weber2005}.

In the collisional QLCA approach, we use the collisional dynamical matrix, whose elements, denoted by $A,B$, are
\begin{equation}
G_{AB}(\vb k,\omega) =  - i \omega R_{AB} + C_{AB}(\vb k).
\label{GAB}
\end{equation}
The elements $C_{AB}(\vb k)$ are those of the QLCA dynamical matrix for a general two-component system \cite{Kalman1990,Golden2000}
\begin{eqnarray}
C^{\alpha\beta}_{AB}(\vb k) &=& -\int \frac{d^3r}{4\pi}  \bigg \{ \omega_{AB}^2\psi^{\alpha\beta}(r)\left [ 1 + h_{AB}(r)  \right ] e^{-i\vb {k \cdot r} }  \nonumber \\
& - & \left. \delta_{AB} \sum_{C}  \Omega_{AC}^2 \psi^{\alpha\beta} (r) \left [ 1 + h_{AC}(r)  \right ] \right \} \nonumber \\
& + &  \delta_{AB} \delta_{\alpha\beta} \sum_{C} \frac{1}{3} \Omega_{AC}^2,
\label{QLCAC}
\end{eqnarray}
\begin{equation}
\psi^{\alpha\beta}(r) = \frac{\partial^2}{\partial r^\alpha \partial r^\beta}\frac{1}{r} = \frac{1}{r^3}\left( 3 \frac{r^\alpha r^\beta}{r^2} - \delta_{\alpha\beta}  \right )   - \frac{4\pi}3 \delta^{\alpha\beta} \delta(\vb r).
\label{psimn}
\end{equation}
The elements $R_{AB}$ represents the damping contribution given by the matrix
\begin{equation}
{\bf R} = \nu \begin{pmatrix}
c_2 & -\sqrt{c_1c_2} \\
-\sqrt{c_1c_2} & c_1
\end{pmatrix}.
\label{RAB}
\end{equation}
Here $\nu = \gamma n/2m$ is a nominal collisional frequency determined solely by the relevant two-body dynamics. The indices $A,B,C$ designate the spin species, $\alpha,\beta$ the Cartesian coordinates, $h_{AB}(r)$ is the pair correlation function between particles in species $A$ and $B$ (note that the two-particle distribution function is $g_{AB}(r) = 1 + h_{AB}(r)$), $\omega_{AB}^2 = \frac{4\pi e^2Z_AZ_B\sqrt{n_An_B} }{\sqrt{m_Am_B}}, \Omega_{AC}^2 = \frac{4\pi e^2 Z_AZ_Cn_C}{m_A}
$ are the nominal plasma and Einstein frequencies, respectively, and $Z_A,m_A$ the charge number and mass of species $A$ with concentration $c_A = n_A/n$. While this derivation is based on classical dynamics, it is expected, as made plausible above, that in the quasi-localized phase of the HEG, it provides a reliable description of the system. It also should be noted that the $h_{AB}(r)$ correlation functions represents an independent input and they should be obtained through the correct quantum dynamics.

Here we are interested in the longitudinal modes only. In particular, we consider their behavior at long wavelengths where the salient features of the mode structures come into focus. Accordingly, for a system of equal masses and charges, the longitudinal projection of eq.~\eqref{QLCAC} gives the longitudinal dynamical matrix elements as
\begin{equation}
C_{\uparrow\uparrow}(k\rightarrow 0) = \omega_0^2 \left ( c_\uparrow - \frac{c_\downarrow}{3}h_{\uparrow\downarrow}(0) + \frac{2}{15}k^2c_{\uparrow}I_{\uparrow\uparrow} \right ),
\end{equation}
\begin{equation}
C_{\downarrow\downarrow}(k\rightarrow 0) = \omega_0^2 \left ( c_\downarrow - \frac{c_\uparrow}{3}h_{\uparrow\downarrow}(0) + \frac{2}{15}k^2c_{\downarrow}I_{\downarrow\downarrow} \right ),
\end{equation}
\begin{equation}
C_{\uparrow\downarrow}(k \rightarrow 0 ) = \omega^2_0 \sqrt{c_\uparrow c_\downarrow} \left ( 1 + \frac13 h_{\uparrow\downarrow} (0) + \frac2{15}k^2 I_{\uparrow\downarrow} \right),
\end{equation}
where $I_{AB} = \int dr \, r h_{AB}(r)$ is proportional to the exchange-correlation potential energy density. The crucial feature of the expressions above is the appearance of the $h_{AB}(0)$ terms, \textit{i.e.} the values of the pair correlation functions at $r=0$. This, in turn, is the consequence of the singular term $\left (4\pi/3 \right ) \delta^{\mu\nu} \delta(\vb r) \left [ 1 + h_{AB}(r)\right ]$ in eq.~\eqref{psimn}. This term, has the trivial values $h_{\uparrow\downarrow}(0) = 0$ in the weak coupling RPA, and $h_{\uparrow\downarrow}(0) =-1$ in the classical BIM. It also plays an important role in atomic systems, where the wave-function and the two-particle function $g(r)$ have a non-vanishing value at $r=0$ \cite{Bethe1977}. Here, it is the crucial element that drives the out-of-phase oscillation of the system.

The complex oscillation frequencies are the solutions of the dispersion relation $||\omega^2\delta_{AB} - G_{AB}(k)|| = 0$, which in the long wavelength limit and to lowest order in $\nu$ are
\begin{equation}
\Omega_+^2(k) = \omega^2_+(k), \quad \Omega_-^2(k) = \omega_-^2(k) - 2 i \nu\omega_-(k),
\label{complexmodes}
\end{equation}
with the real eigenfrequencies
\begin{equation}
\omega^2_+(k\rightarrow 0) = \omega_0^2 \left [ 1 + U(c_{\uparrow},c_\downarrow) k^2 \right ],
\label{omegaplus}
\end{equation}
\begin{equation}
\omega_-^2 (k\rightarrow 0) = \omega_0^2 \left [ - \frac13 h_{\uparrow\downarrow}(0) + V(c\uparrow,c_\downarrow) k^2 \right ],
\label{omegaminus}
\end{equation}
where $U(c_{\uparrow},c_\downarrow) = \frac{2}{15} \left ( c^2_\uparrow I_{\uparrow\uparrow} + 2 c_\uparrow c_\downarrow I_{\uparrow\downarrow} + c^2_\downarrow I_{\downarrow\downarrow} \right )$ and $V(c_{\uparrow},c_\downarrow) = \frac2{15} c_\uparrow c_\downarrow  \left ( I_{\uparrow\uparrow} - 2 I_{\uparrow\downarrow} + I_{\downarrow\downarrow} \right )
$. Eq.~\eqref{omegaplus} represents the well-known plasmon excitation. The new feature is the emergence of the OPO as a dynamical excitation $\omega_-^2$ of the system, eq.~\eqref{omegaminus}, wholly maintained, as expected, by the anti-parallel spin pair correlation function, $h_{\uparrow\downarrow}(r)$ evaluated at $r = 0$. $\omega_-$ can be identified with the characteristic frequency that emerges from the 3\textsuperscript{rd} frequency magnetic sum rule \cite{Goodman1973,Utsumi1983,DAmico2000,Atwal2003,Giuliani2008}. This is the main statement of the present work. 

The expression for $\omega_-$ exhibits the seemingly paradoxical independence of $c_{\uparrow}$ and $c_{\downarrow}$, showing that it does not vanish even when one of the concentration does. It should be realized, however, that it is not the frequency, but the spectral weight of the mode that determines its persistence. It will be shown below eq.~\eqref{spectralweight}, that this latter quantity has the expected behavior indeed. As to the polarized vs. unpolarized ground state the plasmon frequencies at $k=0$ have no explicit dependence on the degree of polarization of the electron liquid (although there is an implicit dependence through $h_{\uparrow\downarrow}(r)$ \cite{Ortiz1994,Ortiz1997,Spink2013}, see also \cite{Gori-Giorgi2002c}); however, the $k$-dependent coefficients do. The $k$-dependence, to lowest order in $k$, can be determined through the $I_{AB}$ integrals, which require the input of all three spin-resolved correlation functions. Such spin-resolved pair correlation function data were generated from diffusion Monte Carlo simulations and are available for the unpolarized HEG up to $r_s = 20$ and for the polarized HEG at $r_s = 3$ (see Fig.~2,3 of Ref.\cite{Spink2013} or Ref.~\cite{Gori-Giorgi2000,Gori-Giorgi2002} for $r_s = 0-10 $). These simulation appear to share one common feature: $h_{\uparrow\downarrow}, h_{\downarrow\downarrow} < 0$, whence $I_{\uparrow\downarrow}, I_{\downarrow\downarrow} < 0$. Thus, the dispersion of the OPO eq.~\eqref{omegaminus} appears to be negative. This is, however, true only to the extent that the positive BG contribution is negligible (\textit{i.e.} the $O(k^4)$ term in eq.~\eqref{w3sumrule}).
A typical value for $h_{\uparrow\downarrow}(0)$ is very close to $-1$ in the strong coupling regime, thus positioning $\omega_-$ at about 57\% of the plasma frequency. The remarkable relationship between $\omega_-$ and the Einstein frequency of the system has already been observed by Goodman and Sj\"{o}lander \cite{Goodman1973}. Here, we note that the existence of this link follows from the general formalism of binary Coulomb systems \cite{Kalman2014}. It may be observed that the lowest value of $\omega_-$, seems to occur in the weak coupling limit when $h_{\uparrow\downarrow}\rightarrow 0$ as $r_s \rightarrow 0$. This is, however, misleading, since in this limit the electrons are weakly coupled and rather than the QLCA it is the RPA that is applicable. The opposite limit, where $h_{\uparrow\downarrow} = -1$, \textit{i.e.} $g_{\uparrow\downarrow} = 0$ is similar to the classical BIM. We observe that the normal plasmon, $\Omega_+$, where the two spin components oscillate in-phase is unaffected (it is still slightly damped by Landau damping, which is not covered by the QLCA model). 

We now turn to comparing the spectral weights of the two plasmon excitations. To do this one needs the collective contribution to the partial DSF-s, $S_{AB}(\vb k,\omega)\propto \braket{n_A(\vb k,\omega)n_B(\vb k,\omega)}/\sqrt{n_An_B}$, which are determined by the imaginary part of the  density-density response function, $\chi''_{AB}(\vb k,\omega) \equiv \IM{\chi_{AB}( \vb k, \omega)}$, whose elements, calculated with the aid of the QLCA formalism, are
\begin{equation}
\chi''_{AB}(\vb k,\omega) =  \frac{ \sqrt{n_A n_B}k^2}m \IM{\left [ \omega^2 \vb I - \vb G( \vb k,\omega) \right ]^{-1}_{AB} }.
\end{equation}
The Fluctuation-Dissipation theorem for the partial DSF-s of the spin-up/spin-down density fluctuation at $T = 0$ provides the collective contribution to $S_{AB}(\vb k,\omega)$, (we restrict the calculation to the vicinity of the OPO):
\begin{equation}
S_{AB}(\vb k,\omega) = - \frac{\hbar}{\pi \sqrt{n_An_B}} \chi''_{AB}( \vb k,\omega).
\end{equation}
Then, a rather lengthy algebra yields, to lowest order in $\nu$ and $k$, a Lorentzian distribution over $\omega^2$, centered around $\omega_-^2$
\begin{equation}
S_{AB}(\vb k,\omega) = \frac{\hbar k^2}{\pi m}\frac{\omega R_{AB} }{\left(\omega^2 - \omega^2_-\right )^2 + \omega^2(R_{\uparrow\uparrow} + R_{\downarrow\downarrow})^2}.
\end{equation}
In view of eq.~\eqref{RAB}, all the elements of $S_{AB}$ are proportional to $c_\uparrow c_\downarrow$ and
$ S_{\uparrow\uparrow} = S_{\downarrow\downarrow} = - S_{\uparrow\downarrow}$.
Consequently, the DSF of the total density fluctuation vanishes $S_{+} = c_{\uparrow} S_{\uparrow\uparrow} + c_{\downarrow} S_{\downarrow\downarrow} + 2 \sqrt{c_{\uparrow}c_{\downarrow} } S_{\uparrow\downarrow} = 0$. This is expected, since the total density remains unaffected by the out-of-phase oscillations of the spin densities. In contrast, all spin-resolved density fluctuations and the magnetic DSF-s, 
$ S_{-} = c_{\uparrow}c_{\downarrow} \left ( c_{\downarrow} S_{\uparrow\uparrow} + c_{\uparrow}S_{\downarrow\downarrow} - 2 \sqrt{c_{\uparrow}c_{\downarrow}} S_{\uparrow\downarrow} \right ) $ exhibit a well-defined peak at $\omega_-$
\begin{equation}
	S_{\uparrow\uparrow} (\vb k,\omega_-) = - S_{\uparrow\downarrow}(\vb k,\omega_-) = \frac{\hbar k^2}{\pi m} \frac{c_\uparrow c_	\downarrow}{ \nu \omega_-}.
	\label{spectralweight}
\end{equation}
We note that $S_-(\vb k,\omega)$ has its maximum in the unpolarized state and vanishes in the fully polarized ferromagnetic phase. Physically $\nu$ is related to the spin drag of the HEG \cite{Weber2005,DAmico2010,Takahashi2007,Yashenkin2015}. In the absence of more physical information on the relevant collision frequency, the important point is that the strength has a finite value, accessible to observation. 

The existence of the OPO mode as part of the collective excitation spectrum should have a bearing on the equilibrium properties of the HEG. It also must be accommodated by the strict alliance of sum rules that guard the totality of excitations in the system. To see how this requirement can be satisfied we analyze the full dynamics of the DSF-s by invoking the Feynman Ansatz (FA) \cite{Feynman1954,Feynman1955,Feynman1956,Cohen1957}, in a form modified to accommodate the features of the current scenario. The original FA is based on the assumptions 
\begin{itemize*}\item[\textit{(i)}] that at $k=0$ the collective mode contribution to the DSF assumes a delta function singularity, and \item[\textit{(ii)}] that there is no significant non-resonant contribution to the DSF.
\end{itemize*}
In the binary system, however, there is damping at $k=0$ caused by the interspecies viscous drag, and there is a non-resonant low frequency addition to the spin-resolved DSF due to exchange \cite{Gori-Giorgi2002b}. In order to account for these features we represent $S_+(\vb k,\omega)$ and $S_-(\vb k,\omega)$, the density and magnetic DSF-s, as follows (only the unpolarized state is covered in the sequel): 
\begin{equation}
S_+(\vb k,\omega) = A_+(k) f_\tau(\omega - \omega_+),
\label{SPlus}
\end{equation}
\begin{equation}
S_-(\vb k,\omega) = BS_0(\vb k,\omega) + A_-(k) f_\sigma(\omega - \omega_-).
\label{SMinus}
\end{equation}
Here $f_{\tau,(\sigma)}(\omega - \omega_{+,(-)})$ are normalized distributions, possessing finite frequency moments, characterized by width $\tau (\sigma)$ ($f_0(x) = \delta(x)$), $ A_\pm (k) = \pi a_\pm k^2$, and
\begin{equation}
S_0(\vb k,\omega) = \sum_{p} \delta \left (\omega - \omega_{\vb{pk} }\right ) n_{\vb p}\left [ 1 - n_{\vb {p+k}}\right ]
\end{equation}
is the exchange contribution with $\omega_{\vb {pk}} = \varepsilon_{\vb {p + k}} - \varepsilon_{\vb p}$ and $\varepsilon_{\vb p} = |\vb p|^2/2m$. $S_0(\vb k,\omega)$ is calculated via the unperturbed Fermi distribution (even though it is the actual $n_{\vb p}$ correlational distribution that should be used here, it is expected that the error resulting from this replacement in $S_0(\vb k,\omega)$ would be canceled by other higher order terms \cite{Ziesche2010,Atwal2003}). Calculating the $p$-th sum rule for $p = 0,1,3$ using eq.~\eqref{SMinus} we obtain 
\begin{equation}
	\braket{\omega^0} \rightarrow \frac{3}4B k + a_-k^2 =  S_-(k)
\end{equation}
\begin{equation}
	\braket{\omega^1} \rightarrow  B \frac{\hbar k^2}{m} + a_- \omega_- \xi_\sigma k^2  = \frac{\hbar k^2}{m}  
\end{equation}
\begin{equation}
\braket{\omega^3} \rightarrow O(k^4) +  a_-\omega_-^3\eta_\sigma k^2  = \frac{\hbar k^2}{m} \left [ C_{-}(\vb k) + O(k^2) \right ]
\label{w3sumrule}
\end{equation}
where $C_-(\vb k) = C_{\uparrow\uparrow}(\vb k) - C_{\uparrow\downarrow}(\vb k)$ and $C_{AB}(\vb k)$ are calculated using eq.~\eqref{QLCAC}, and $\xi_\sigma,\eta_\sigma$ are algebraically determined functions of $\sigma$. The $\braket{\omega^3}$ sum rule is satisfied for $\omega_-$ given by eq.~\eqref{omegaminus}. For given values of  $\nu$ and $f_\sigma(0)$ we can determine the width $\sigma$ from which we calculate the function $\eta_\sigma,\xi_\sigma$ and obtain the amplitude $B = 1- \xi_\sigma/\eta_\sigma$ (necessarily $<1$). With more information on the spin drag coefficient the development of a numerically more explicit model with more predictive value should become possible.

While the dynamical binary model presented here is an inherently classical one, there is little doubt that the same physics applies to a degenerate electron liquid. The main new effect that enters the mechanism of generation of collective modes stems from the required anti-symmetrization of the localized wave functions. However, the only modifications of the above results show up mostly at finite $k$-values. This is not surprising since exchange is a short-range effect. There is no modification over the classical result at $k=0$.  

Finally we wish to relate to a line of research on a related collective excitation that has gained importance recently and goes under the name of ``spin-plasmon" (SP) \cite{Agarwal2014}. The similarity between the OPO and the SP is that both modes represent out-of-phase oscillations of the up and down spin densities. In contrast to the OPO, the SP exists only in the strongly polarized state of the HEG, when there is a substantial density difference between the majority and minority spin populations. It is the ensuing separation of the respective Fermi velocities that makes the excitation of the SP possible. Thus, the SP is not maintained by correlations and therefore it may exist in the weakly coupled phase and can be described within the RPA formalism. The mechanism of the SP is very similar to the one responsible for the existence of the ion acoustic mode in conventional plasmas with disparate electron and ion temperatures: here the difference of the Fermi energies replaces the disparity of the temperatures. For these reasons the frequency of the SP is determined by the Fermi velocities of the components and is independent of the correlation function and thus it is quite different from the OPO frequency. It also should be noted that the SP mechanism has been worked out for the 2D HEG only: although it is plausible to assume that the 3D scenario would not be substantially altered, the different behaviors of the 2D and 3D Lindhard functions may play an unexpected role. How the OPO and the SP would relate to each other in a highly polarized and strongly coupled system is a question of great interest that would deserve further study.

In summary, we have shown the existence of a second plasmon $\omega_-$, the OPO, in a strongly coupled electron liquid. The OPO is maintained by the out-of-phase oscillations of the spin-up and spin-down components of the electron liquid, but governed solely by the Coulomb interaction between the particles. However, it is the difference between the parallel and anti-parallel spin density correlation functions which is responsible for $\omega_-$ being proportional to the overlap ($r=0$) (absolute) value of $h_{\uparrow\downarrow}(0)$, which is an increasing function of the coupling parameter, $r_s$. The spectral weight of the mode is proportional to the product of the spin-up and spin-down densities: thus it vanishes in the completely polarized state. It has a typical Lorentzian structure, with a peak value which, again, depends on $h_{\uparrow\downarrow}(0)$. However, since the total density is not influenced by the relative oscillations of the two components, the spectral weight is exhibited in the magnetic DSF only, leaving the total density structure function unaffected.

There are several physical ingredients that contribute to the formation of this collective excitation. First, an interacting many-particle system may always sustain relative oscillations between any, arbitrarily chosen group of particles. Second, in the HEG the electron spin can serve as an identifiable and observable marker of the two components. Third, the fact that the spin has only two quantum states, makes the system for the purpose of the physical scenario involved here isomorphic to a classical binary liquid. Fourth, in contrast to classical systems, the quantum behavior makes it possible for the two-particle wave-function to assume a non-vanishing value at zero separation. Fifth, the strong coupling between the electrons leads to their quasi-localization. None of these relates to spin-spin interaction, ensuring that the OPO is an excitation independent from conventional spin waves.

We have estimated the range of $r_s$ values within which the OPO manifests itself as $r_s = 10-60$. Within this range the electron liquid may become polarized: this has no bearing on the $k = 0$ frequency of the mode, but it affects its spectral weight with a tendency to quench it. We estimate the typical value of $\omega_-$ to be about 57\% of the normal plasmon frequency. The $k$-dependent dispersion of $\omega_-(k)$, similarly to the normal plasmon dispersion, is the combination of a positive BG contribution governed by the kinetic energy (with a value that can be determined by invoking a two-component third moment sum rule) and a negative contribution, governed by the spin-spin exchange-correlation energies. In a polarized state, both of these factors are affected by the degree of polarization. The damping of the mode in the long wavelength limit is expected to be governed by electron-electron collision induced spin drag~\cite{DAmico2010,Takahashi2007,Yashenkin2015,Weber2005}. 
We conclude with a note on the case of finite temperature. The presence of the OPO excitation persists even in these conditions, albeit with a diminished spectral weight. Our formalism, as presented, does not account for thermal fluctuations; therefore, our estimation of the OPO's manifestation is primarily focused on high $r_s$ values. Notably, recent advancements in computational techniques have enabled the calculation of the dynamic structure factor, $S_+(k,\omega)$, for the homogeneous electron liquid in the warm dense matter regime~\cite{Dornheim2023}.It is important to recognize that warm dense matter typically exhibits $r_s$ values of around unity. We are optimistic that continued progress in computational methods will facilitate dynamic simulations in regimes of stronger $r_s$ values, where the OPO excitation's presence is expected to be more pronounced.

\acknowledgements
This work has been partially supported by NSF Grant PHY-0715227 and PHY-1105005. Useful discussions with Pradip Bakshi, Kevin Bedell, Gaetano Senatore, and Alessandro Principi are gratefully acknowledged.

This paper represents the final collaborative effort of its authors, G.J.K., K.I.G., and L.G.S. The original idea was conceived by G.J.K. and K.I.G., whose insights laid the groundwork for this work. L.G.S. performed the calculations actualizing their collective vision. The original manuscript was drafted by all three authors before the passing of K.I.G. and G.J.K. In honor of the invaluable advice and mentorship L.G.S. received from G.J.K., L.G.S. took on the task of completing the manuscript. His commitment to this work is both a professional dedication and a personal tribute to his late advisor and colleague.
Keeping G.J.K. and K.I.G. as co-authors posthumously recognizes their essential contributions and pays homage to their enduring legacy in the field. L.G.S.'s completion of the paper, rooted in deep gratitude and respect for his mentors, highlights the lasting impact of their collaboration. Thus, this paper is not just a significant academic contribution but also a symbol of respect and remembrance for G.J.K. and K.I.G., whose influence resonates through this final shared accomplishment.

\bibliography{References}
\bibliographystyle{abbrv}
\end{document}